-------------------------------------------------------------------------
---------------------------------------------------------------------------
\documentstyle[12pt]{article}

\newcommand{\EQ}{\begin{equation}}
\newcommand{\EN}{\end{equation}}

\begin{document}

\topmargin 0pt
\oddsidemargin 5mm
\newcommand{\NP}[1]{Nucl.\ Phys.\ {\bf #1}}
\newcommand{\PL}[1]{Phys.\ Lett.\ {\bf #1}}
\newcommand{\NC}[1]{Nuovo Cimento {\bf #1}}
\newcommand{\CMP}[1]{Comm.\ Math.\ Phys.\ {\bf #1}}
\newcommand{\PR}[1]{Phys.\ Rev.\ {\bf #1}}
\newcommand{\PRL}[1]{Phys.\ Rev.\ Lett.\ {\bf #1}}
\newcommand{\MPL}[1]{Mod.\ Phys.\ Lett.\ {\bf #1}}
\newcommand{\JETP}[1]{Sov.\ Phys.\ JETP {\bf #1}}
\newcommand{\TMP}[1]{Teor.\ Mat.\ Fiz.\ {\bf #1}}

\renewcommand{\thefootnote}{\fnsymbol{footnote}}

\newpage
\setcounter{page}{0}
\begin{titlepage}
\begin{flushright}
SISSA-EP-154
\end{flushright}
\vspace{0.5cm}
\begin{center}
{\large  RG flows and resonance scattering amplitudes
} \footnote{To appear in the Proceedings of the CAP/NSERC workshop on
``Quantum Groups, Integrable Models and Statistical Systems''(July 1992),
J. LeTourneux, L. Vinet (eds), World Scientific.} \\
\vspace{1cm}
\vspace{1cm}
{\large  M\'arcio Jos\'e  Martins
\footnote{on leave from Departamento de Fisica, Universidade Federal de
S.Carlos, C.P. 676 - S.Carlos 13560, Brazil.}
\footnote{martins@itssissa.bitnet}} \\
\vspace{1cm}
{\em International School for Advanced Studies\\
34014, Strada Costiera 11, Trieste,
 Italy}\\
\end{center}
\vspace{1.2cm}

\begin{abstract}
We review recent progresses in the study of factorized
resonance scattering S-matrices. The resonance amplitudes
are introduced through a suitable analytical continuation
of the ADE Toda S-matrices. By using the
thermodynamic Bethe ansatz approach we
are able to compute the
ground state energy, which describes a rich
pattern of flows interpolating between the central
charges of the coset models based on the ADE
Lie algebras. We also present the simplest resonance
``$\phi^3$'' scattering model and discuss its relation
with new flows in non-unitary minimal models. Further
generalizations are discussed in terms of certain asymptotic
conditions in a family of ``resonance'' functional
hierarchies.
\end{abstract}
\vspace{.2cm}
\vspace{.2cm}
\centerline{August 1992}
\vspace{.25cm}
\end{titlepage}

\renewcommand{\thefootnote}{\arabic{footnote}}
\setcounter{footnote}{0}

\newpage
\section{Introduction}
The study of scale non-invariant systems is certainly interesting
and in many cases a difficult open problem. A relevant aspect
of this problem is, e.g., the study of the renormalization group~(RG)
scenario in the vicinity of the fixed point. Not much is known
about this problem in higher space time dimension. Recently, however,
in (1+1) dimension a considerable progress has been achieved
after the work of Belavin, Polykov and Zamolodchikov \cite{BPZ}.
Considering a unitary theory, the RG scenario is severely constrained
by Zamolodchikov's c-theorem\cite{ZAM1}. The
c-theorem tells us that RG flows
always run down hill, thinning the degrees of freedom as the
RG trajectory flows from
the ultraviolet region to the infrared regime. A.B. Zamolodchikov \cite{ZAM2}
has considered a non-scale invariant field theory as a relevant
perturbation of the conformal field theory which characterizes the
ultraviolet properties of the system. In certain cases the theory is
integrable and factorizable S-matrices can be conjectured \cite{ZAM2}.
Using the S-matrices one can, in principle, apply the thermodynamic
Bethe ansatz \cite{YY} (TBA) approach in order to determine the
ground state scaling function \cite{Al1,Al2,Al3} associated to the
respective RG trajectories. A typical example is the minimal models
$M_p$ perturbed by the least relevant operator $\phi_{1,3}$. The
infrared behaviour is highly dependent of the sign of the perturbation:
it will induce a crossover either to the lower critical theory
$M_{p-1}$ \cite{ZAM3,LUCA} or to a purely massive theory of (p-2) types
of Kink-antiKink pairs \cite{ZAM4,TUDO}.

The main purpose of this article is to discuss the RG trajectories associated
to the recently discovered resonance
scattering models \cite{Al4,M1,M2,DR}. In section 2
we introduce a general class of resonance S-matrices based on the ADE
affine Toda scattering amplitudes \cite{AFZ,CRMU,BC,DV}. This
section also contains an analysis of the associated Casimir energy
via TBA approach. Finally, we discuss the simplest ``$\phi^3$''
resonance factorized scattering model and its connection with
new flows in non-unitary minimal models. In Section 3 we discuss
the functional hierarchies of the TBA equations
and we also investigate a further generalization of these
relations.

\section{The resonance S-matrix and its Casimir energy}

Recently Al. Zamolodchikov \cite{Al4} has proposed the simplest
resonance S-matrix consisting of a single particle scattering
through the amplitude
\EQ
S(\theta,\theta_0)= \frac{\sinh(\theta)-i \cosh(\theta_0)}
{\sinh(\theta)+i \cosh(\theta_0)}
\EN
where $\theta_0$ is the resonance parameter.

Eq.(1) can be obtained from the sinh-Gordon S-matrix through a
suitable analytical continuation of its coupling constant. This fact
suggests that a more general class of resonance S-matrices can be
formulated from the ADE Toda theory \cite{AFZ,CRMU,BC,DV}. Indeed,
the amplitude $S_{a,b}(\theta,\theta_0)$ of the resonance
ADE S-matrices \cite{M1,M2,DR} can be defined by
\EQ
S_{a,b}(\theta,\theta_0)=S_{a,b}^{min}(\theta)
Z_{a,b}(\theta,b(\alpha)=\frac{\pi}{h} \pm i \theta_0)
\EN
where $S_{a,b}^{min}(\theta)$ are the minimal amplitudes containing
the physical poles; $Z_{a,b}(\theta,b(\alpha))$ are the so-called
Z-factors which encode the coupling constant $\alpha$ through
the function $b(\alpha)$ and $h$ is the Coxeter number. It is
easy to check that Eq.(2) reproduces Al. Zamolodchikov model
for the $A_1$ Lie algebra.

Our interest now is to study the finite size corrections of
the ground state associated to these resonance S-matrices. The
TBA equations \cite{Al1,Al2,Al3} describe exactly the ground
state energy of an integrable theory on a torus of radius R.
In our case the Casimir energy $E(R,\theta_0)$ is given by
\EQ
E(R,\theta_0) = -\frac{1}{2 \pi}
\sum_{a=1}^{r} m_a \int_{-\infty}^{+\infty} d\theta \cosh(\theta)
L_a(\theta)
\EN
where $m_a$ are the ADE mass gaps \cite{AFZ,CRMU,BC,DV},
$r$ is the rank of the respective Lie algebra, and
$L_a(\theta)=\ln(1+e^{-\epsilon_a(\theta)})$. The pseudoenergies
$\epsilon_a(\theta)$ satisfy the following integral equation~(TBA
equations)
\EQ
\epsilon_a(\theta) + \frac{1}{2\pi} \sum_{b=1}^{r}
\psi_{a,b}*L_b(\theta) =m_a R \cosh(\theta)
\EN
where $\psi_{a,b}(\theta)=-i \frac{d}{d \theta} S_{a,b}(\theta,\theta_0)$
and the symbol $f*g(x)$ denote the convolution
$f*g(x)=
\frac{1}{2 \pi}
\int_{-\infty}^{\infty}f(x-y)g(y) dy$.

The ultraviolet limit of Eq.(1) is dominated by a background conformal
theory with central charge $r$, and the first correction is
logarithmic in R \cite{Al4,M2}. In terms of the function $c(R,\theta_0)=
-\frac{6R}{\pi} E(R,\theta_0)$ our result reads,
\EQ
c(R,\theta_0)= r-\frac{3 (\theta_0^2+\frac{\pi^2}{h^2})}
{X^2} \sum_{a,b} C_{a,b}^{-1}
\EN
where $X=\ln(\frac{m_1 R}{2})$.

In order to analyze the behaviour of function $c(R,\theta_0)$ for
intermediated distances of R we have numerically solved Eq.(4). Figures
1(a,b) show the behaviour of $c(R,\theta_0)$ for $\theta_0=20,40$ in the
case of the $A_2$ Lie algebra. After a careful analysis of other
models, we arrived at the following general picture \cite{Al4,M1,M2}.
At $\theta_0=0$ function $c(R,\theta_0)$ presents a smooth behaviour
between the ultraviolet and infrared regimes. However, for large
enough $\theta_0$, $c(R,\theta_0)$ starts to form plateaux around
the central charges $c_p^r= r(1-h(h+1)/p(p+1)), p=h+1,h+2,...$ of the
$G_1 \otimes G_{p-h}/ G_{p-h+1}$ models. More precisely, each time
that $X \simeq -(p-h) \frac{\theta_0}{2}$ function $c(R,\theta_0)$
crosses over from its value $c_p^r$ to the next (up) value $c_{p+1}^r$.
It is important to stress here that the ``RG time'' $\frac{\theta_0}{2}$
accounts for the plateau and its finite size corrections~(see fig.1(a,b)).
Discussions on the identification of this ``staircase pattern'' with
a deformed conformal field theory can be found in refs. \cite{LA,Al4,M2}.

Let us now introduce what we believe to be the simplest resonance
scattering model possessing the ``$\phi^3$''-property. The model consists
of a single particle $a$ and its two-body S-matrix is given by
\EQ
S_{a,a}(\theta,\theta_0)=\frac{\tanh\frac{1}{2}(\theta+i\frac{2 \pi}{3})}
{\tanh\frac{1}{2}(\theta-i\frac{2 \pi}{3})}
\frac{\tanh\frac{1}{2}(\theta-\theta_0-i\frac{\pi}{3})}
{\tanh\frac{1}{2}(\theta-\theta_0+i\frac{\pi}{3})}
\frac{\tanh\frac{1}{2}(\theta+\theta_0-i\frac{\pi}{3})}
{\tanh\frac{1}{2}(\theta+\theta_0+i\frac{\pi}{3})}
\EN

The Toda related field theory is the one analyzed by Arinshtein et al
\cite{AFZ} and known as the Shabat-Mikhailov model. The ``minimal''
part of Eq.(6) is the S-matrix \cite{CM} of the perturbed Yang-Lee
edge singularity \cite{FC}. We also notice that amplitude
$S_{a,a}(\theta,\theta_0)$ satisfy the following important
relation
\EQ
S_{a,a}(\theta,\theta_0)= S_{1,1}(\theta,\theta_0)S_{1,2}(\theta,\theta_0)
\EN
where $S_{1,1}(\theta,\theta_0)$ and $S_{1,2}(\theta,\theta_0)$
are the $A_2$ amplitudes.

{}From Eqs.(7) and (4) it follows
that the ground state energy of this theory is precisely half of that
of the $A_2$ model. Hence, the plateau will now form around the
values $c_p=1-12/p(p+1), p=4,5,...$. Taking into account that the effective
central charge of the minimal models $M_{\frac{p}{q}}$ is
$c_{ef}=1-6/pq$, we are able to identify the following non-unitary
minimal models:
$M_{\frac{q}{2q+1}}$~(q=p/2, p even) and
$M_{\frac{q+1}{2q+1}}$~(q=(p-1)/2, p odd). Futhermore, from the
finite size corrections of the ground state we identify the fields
$\phi_{2,1}$ and $\phi_{1,5}$ as those responsible for the crossover
behaviour. More precisely, we predict the following new flows
\EQ
M_{\frac{q+1}{2q+1}} + \phi_{2,1} \rightarrow M_{\frac{q}{2q+1}} ~~~q=2,3,...
\EN
and for $M_{\frac{q}{2q+1}}$,
\EQ
M_{\frac{q}{2q+1}} + \phi_{1,5} \rightarrow M_{\frac{(q-1)+1}{2(q-1)+1}}
{}~~~q=3,4,...
\EN

We remark that the field $\phi_{2,1}$ and $\phi_{1,5}$ cannot be
both relevant operators in the same model. In addition, as already
discussed by the author \cite{M3}, the $\phi_{1,5}$ perturbation
can formally be related to the $\phi_{1,2}$ deformation.
We believe that the combination $\lambda \phi_{2,1}
+\tilde{\lambda} \phi_{1,5}$ play a similar role of the
fields
$\lambda \phi_{1,3}
+\tilde{\lambda} \phi_{3,1}$ appearing in the minimal models \cite{LA}.
For other relevant discussions see ref. \cite{M2}.

\section{Resonance functional hierarchies}

In this section we discuss certain functional relations for
functions $Y_a(\theta)=e^{\epsilon_a(\theta)}$. It is possible
to rewrite the TBA equations in a more suggestive way, adopting a
similar approach as in ref. \cite{Al5}. First one has to notice the
following remarkable matrix identity,
\EQ
{\left [ \delta_{a,b} -\frac{\tilde{\psi}_{a,b}(k,\theta_0)}{2 \pi}
\right ]}^{-1}= \frac{\delta_{a,b} \cosh[\frac{\pi k}{h}] -l_{a,b}/2}
{\cosh[\frac{\pi k}{h}] -\cos(k \theta_0)}
\EN
for the fourier component $\tilde{\psi_{a,b}}(k,\theta_0)=
\int_{-\infty}^{+\infty} e^{i k \theta} \psi_{a,b}(\theta,\theta_0) d \theta$.
Here,
 $l_{a,b}$ is the incident matrix  of the
G=A,D,E Lie algebra.

Using Eq.(10) and the relation $m_a=\sum_{b=1}^{r} l_{a,b} m_b$, we obtain
the set of functional equations given by
$Y_a(\theta)=e^{\epsilon_a(\theta)}$ \cite{Al4,M2},
\EQ
Y_a(\theta +\frac{i \pi}{h}) Y_a(\theta-\frac{i \pi}{h})=
\prod_{b \in G} {\left [1+Y_b(\theta) \right ]}^{l_{a,b}}
{ \left [1+Y_a^{-1}(\theta+\theta_0) \right ] }^{-1}
{ \left [1+Y_a^{-1}(\theta-\theta_0) \right ] }^{-1}
\EN

We stress that one may start with such functional hierarchies instead
of considering the ADE resonance S-matrices of section 2. The connection
with the TBA equations is due to the asymptotic conditions of the
functions $Y_a(\theta)$ at low temperature, i.e., $Y_a(\theta)
\simeq exp(m_a R \cosh(\theta))$ $R \rightarrow \infty$. A simple
generalization of Eq.(11) is the one discussed by the author in
ref. \cite{M4}. We have two pairs of functions $Y_a^1(\theta)$ and
$Y_a^2(\theta)$ satisfying
\begin{eqnarray}
Y_a^1(\theta +\frac{i \pi}{h}) Y_a^1(\theta-\frac{i \pi}{h})  =
\prod_{b \in G} {\left [1+Y_b^1(\theta) \right ]}^{l_{a,b}}
{ \left [1+1/Y_a^2(\theta+\theta_0) \right ] }^{-1}
{ \left [1+1/Y_a^2(\theta-\theta_0) \right ] }^{-1} & & \nonumber \\
Y_a^2(\theta +\frac{i \pi}{h}) Y_a^2(\theta-\frac{i \pi}{h})  =
\prod_{b \in G} {\left [1+Y_b^2(\theta) \right ]}^{l_{a,b}}
{ \left [1+1/Y_a^1(\theta+\theta_0) \right ] }^{-1}
{ \left [1+1/Y_a^1(\theta-\theta_0) \right ] }^{-1} & &
\end{eqnarray}

Considering the non-trivial compatible asymptotic conditions
$Y_a^1(\theta) \simeq exp(m_a R \cosh(\theta))$ and $
Y_a^2(\theta) \simeq 1$, we found the following consistent
TBA equations \cite{M4}
\begin{eqnarray}
\epsilon_a^1(\theta) +  \sum_{b=1}^{r}
\phi_{a,b}*L_b^1(\theta)+\sum_{b=1}^{r} \varphi_{a,b}*L_b^2(\theta) & = &
m_a R \cosh(\theta) \nonumber \\
\epsilon_a^2(\theta) +  \sum_{b=1}^{r}
\phi_{a,b}*L_b^2(\theta)+\sum_{b=1}^{r} \varphi_{a,b}*L_b^1(\theta) & = & 0
\end{eqnarray}
where $L_a^i(\theta)=\ln(1+e^{-\epsilon_a^i(\theta)}), i=1,2$.
Functions $\phi_{a,b}(\theta)$
and $\varphi_{a,b}(\theta)$ are related to
the ADE Toda scattering amplitudes \cite{AFZ,CRMU,BC,DV}
through the relations
$\phi_{a,b}(\theta)=-i\frac{d}{d \theta}\ln S_{a,b}^{min}(\theta)$ and
$\varphi_{a,b}(\theta)=-i\frac{d}{d \theta}
\ln Z_{a,b}(\theta,b=\frac{\pi}{h} \pm i \theta_0)$.

{}From our analysis of the respective Casimir energy, function
$c(R,\theta)$ interpolates between the central charges of the
$G_2 \otimes G_l /G_{l+2}$~(l even) coset models. The ``RG time'' being
double the amount of the one found in section 2, namely $\theta_0$.
We believe that this observation will
be helpful for further generalizations of our functional
relations. It is also worth to mention that similar functional
relations have been discussed in the literature \cite{PEKU,NA}
in the context of the inversion relations of integrable lattice
models. Such relations play the keystone in the computations of
critical exponents \cite{PEKU,NA} and probably still hide a large amount of
information yet to be explored.

\section*{Acknowledgements}
The author would like to thank the organizers of the workshop ``Quantum
Groups, Integrable Models and Statistical Systems'' for their hard work,
and for providing me an opportunity to present this material.

\newpage
\centerline{\bf Figures}
\vspace{19cm}
Fig. 1(a,b) The scaling function $c(R,\theta_0)$ for the $A_2$ model:
(a)$\theta_0=20$, (b) $\theta_0=40$
\end{document}